\def\year{2022}\relax
\documentclass[letterpaper]{article} 
\usepackage{aaai22}  
\usepackage{times}  
\usepackage{helvet}  
\usepackage{courier}  
\usepackage[hyphens]{url}  
\usepackage{graphicx} 
\usepackage{natbib}  
\usepackage{caption} 

\DeclareCaptionStyle{ruled}{labelfont=normalfont,labelsep=colon,strut=off} 
\usepackage{algorithm}
\usepackage{algorithmic}
\usepackage{cite}
\usepackage{amsmath,amssymb,amsfonts}
\usepackage{latexsym}
\usepackage{pifont}
\usepackage{algorithmic}
\usepackage{graphicx}
\usepackage{textcomp}
\usepackage{xcolor}
\usepackage{newfloat}
\usepackage{listings}
\definecolor{codegreen}{rgb}{0,0.6,0}
\definecolor{codegray}{rgb}{0.5,0.5,0.5}
\definecolor{codepurple}{rgb}{0.58,0,0.82}
\definecolor{backcolour}{rgb}{0.95,0.95,0.92}

\lstdefinestyle{mystyle}{
    backgroundcolor=\color{backcolour},   
    commentstyle=\color{codegreen},
    keywordstyle=\color{magenta},
    numberstyle=\tiny\color{codegray},
    stringstyle=\color{codepurple},
    basicstyle=\ttfamily\footnotesize,
    breakatwhitespace=false,         
    breaklines=true,                 
    captionpos=b,                    
    keepspaces=true,                 
    numbers=left,                    
    numbersep=5pt,                  
    showspaces=false,                
    showstringspaces=false,
    showtabs=false,                  
    tabsize=2
}
\floatstyle{ruled}
\newfloat{listing}{tb}{lst}{}
\floatname{listing}{Listing}
\title{Towards Seamless Management of AI Models in High-Performance Computing}

\author{
Sixing Yu\textsuperscript{\rm 1}, Murali Emani\textsuperscript{\rm 2}, Chunhua Liao\textsuperscript{\rm 3}, Pei-Hung Lin\textsuperscript{\rm 3}, 
Tristan Vanderbruggen\textsuperscript{\rm 3},
Xipeng Shen\textsuperscript{\rm 4}, Ali Jannesari\textsuperscript{\rm 1}
}
\affiliations{
    \textsuperscript{\rm 1}Iowa State University, Ames, IA\\
        \textsuperscript{\rm 2}  Argonne National Laboratory, Lemont, IL\\
       \textsuperscript{\rm 3}  Lawrence Livermore National Laboratory, Livermore, CA\\
       \textsuperscript{\rm 4} North Carolina State University, Raleigh, NC\\
    yusx@iastate.edu, memani@anl.gov, \{liao6, lin32,vanderbrugge1@llnl.gov\}@llnl.gov, xshen5@ncsu.edu, jannesar@iastate.edu

}

\usepackage{bibentry}

\begin{document}

\maketitle

\maketitle

\begin{abstract}

With the increasing prevalence of artificial intelligence~(AI) in diverse science/engineering communities, AI models emerge on an unprecedented scale among various domains. 
However, given the complexity and diversity of the software and hardware environments, reusing AI artifacts (models and datasets) is extremely challenging, especially with AI-driven science applications.
Building an ecosystem to run and reuse AI applications/datasets at scale efficiently becomes increasingly essential for diverse science and engineering and high-performance computing~(HPC) communities.
In this paper, we innovate over an HPC-AI ecosystem – HPCFair, which enables the Findable, Accessible, Interoperable, and Reproducible~(FAIR) principles. 
HPCFair enables the collection of AI models/datasets allowing users to download/upload AI artifacts with authentications. Most importantly, our proposed framework provides user-friendly API for users to  easily run inference jobs 
and 
customize AI artifacts to their tasks as needed.
Our results show that, with HPCFair API, users irrespective of technical expertise in AI, 
can easily leverage AI 
artifacts to their tasks with minimal effort.
\end{abstract}

\section{Introduction}


With the outstanding performance achieved by artificial intelligence~(AI) and machine learning~(ML), AI artifacts~(AI models and datasets) are being increasingly adopted in diverse science and engineering domains, such as materials discovery, ecology, cosmology, biology, and wildlife conservation. However, given the complexity and diversity of the software
and hardware environments, reusing AI artifacts is extremely challenging, especially with AI-driven science and engineering
applications. 
Additionally, AI artifacts developed in 
various scientific domains make it extremely challenging 
for scientists to fetch, reuse, and reproduce. 
Introducing frameworks to reasonably access, reproduce and run those AI applications at scale for diverse science and engineering communities, becomes crucial to accelerate science with 
high-performance computing (HPC).

We first list the key challenges for diverse scientific communities to apply AI artifacts, which need to be addressed by such an AI artifact management framework.
First, AI artifacts rely on complex software and hardware dependencies.
Second, the dependencies vary across AI artifacts. For any given AI artifact, we need to configure running environments for it. 
Third, AI artifacts supported by different backend implementations (e.g., C++ and Python) 
usually have interoperability challenges.
Fourth, applying AI artifacts requires diverse domain scientists'  significant programming skills beyond science.
Fifth, it is hard to find, access, interoperate, and reproduce a target AI model available in public repositories.
Sixth, it is hard for scientists to find a target model that matches their needs perfectly, while customizing AI artifacts need significant efforts, e.g, hundreds of hyper-parameters for tuning
Lastly, there is a lack of benchmark and standardization processes due to which the experimental results are hard to reproduce on the user's customized tasks.

Although the existing HPC-AI artifact management ecosystem~\cite{wolf2019huggingface, chard2019dlhub}  significantly simplifies the threshold for applying AI artifacts, nevertheless, they are dedicated to serving computer science and software engineering domain scientists only.
Such challenges have barely been addressed by existing HPC-AI ecosystems.

In this paper, we propose novel techniques to HPCFair \cite{verma2021hpcfair,nan2021deep} –  an HPC-AI model and data management system, which enables AI artifacts Findable, Accessible, Interoperable, and Reproducible~(FAIR principles) as well as
provides user-friendly interfaces/APIs for diverse domain scientists adopting AI artifacts to their in-demand research tasks.
Specifically, HPCFair containerized AI artifacts, where all the executing dependencies for given artifacts are built in an associate virtual machine independently. Therefore, the proposed work provides users with a friendly executing environment and bypasses the labor-costly environment established on both hardware and software.
Besides that, we designed an HPC ontology to efficiently implement FAIR principles, which enables scientists to easily share and fetch target AI artifacts.

We summarize our contributions as follows:
\begin{itemize}
    \item We proposed a novel AI model knowledge management system for high-performance computing.
    \item Our proposed solution significantly simplified AI model deployment for domain scientists.
    \item It provides user-friendly APIs for scientists to customize AI products to their demands.
\end{itemize}

\begin{figure*}[t]
\begin{center}

\centerline{\includegraphics[width=.8\linewidth]{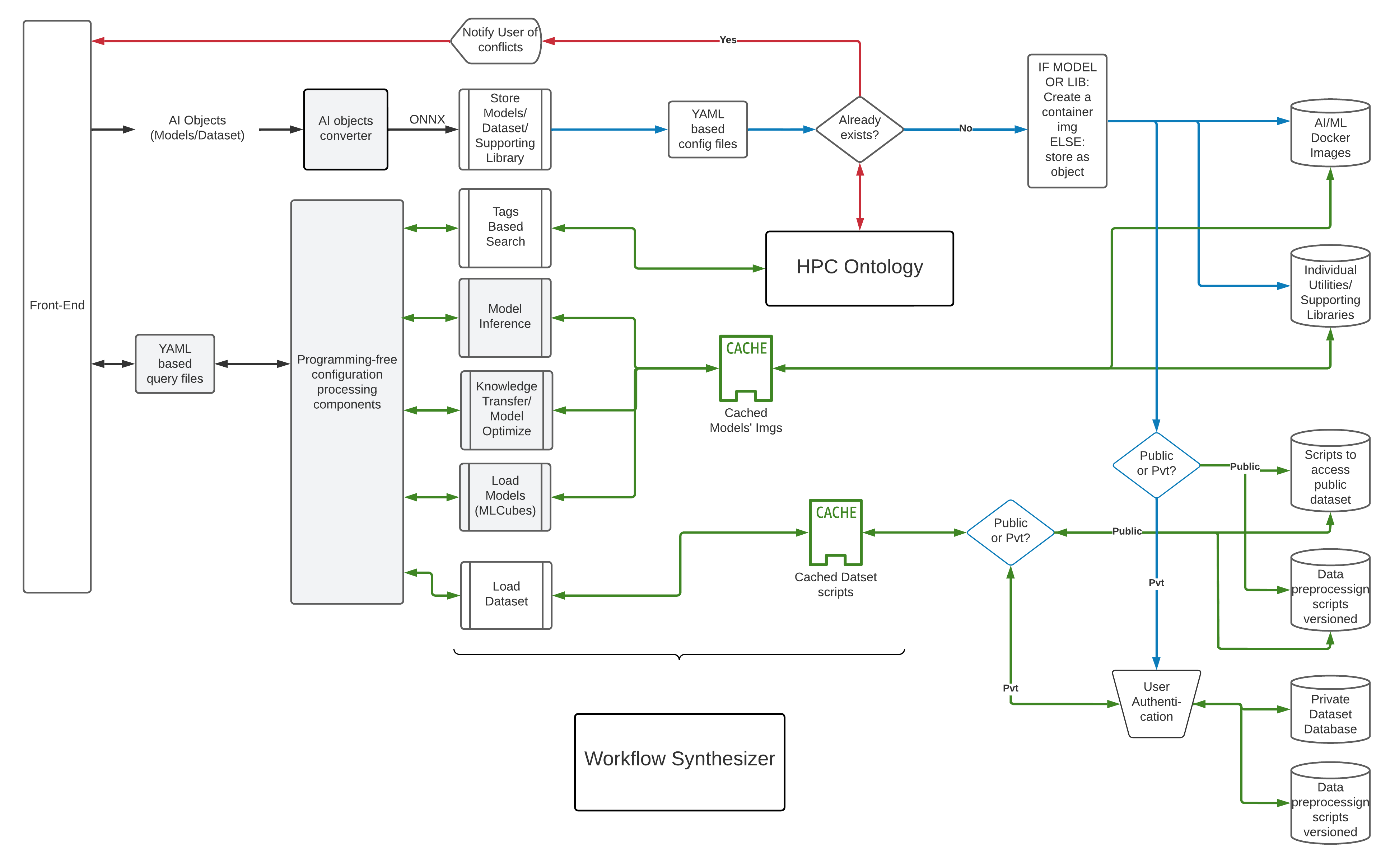}

}
\caption{Designed Workflow for HPCFair.}
 
\label{fig:overview}
\end{center}
\vskip -0.2in
\end{figure*}



\section{Background}
Since AI artifacts popped up on a giant scale, extensive efforts have been devoted to developing efficient AI artifact management tools. In this section, we summarized the State-of-The-Art~(SoTA) AI artifacts tools.

\subsection{Container Platform}
A recent popular trend to improve the reproducibility of AI artifacts is containerization, which enables developers to pack the source code as well as running dependencies and provides an operation system-independent virtual environment for executing target AI artifacts.
SoTA containerized platform such as Docker~\cite{merkel2014docker} and Singularity~\cite{kurtzer2013singularity} enables developers to integrate their codes and dependencies into containers—standardized executable components, and hence, executable in any operating system. 
Nowadays, great efforts are devoted to specializing in containerized machine learning~(ML) models and datasets, such as MLCube~\cite{kahng2016mlcube}.
However, existing containerized platforms are targets to developers publish their works and require expert knowledge for configuration. It is challenging for domain scientists to use in their scientific applications. 

\subsection{AI artifacts Hub}
AI artifacts Hubs gather collections of AI models and datasets and provide a user-friendly interface to search and reproduce AI artifacts.
For instance, Data and Learning Hub for Science (DLHub)~\cite{chard2019dlhub}, a cloud-hosted learning system, enables developers to publish their models with flexible access control.
Collective Knowledge Framework (cKnowledge)~\cite{fursin2021collective} constructed a database of AI components as well as provides APIs and terminal interfaces to efficiently manage research projects for developers. 
Hugging Face~\cite{wolf2019huggingface} offers NLP models and datasets, such as Transformer models, as multi-platform supportive open-source libraries that help users download, infer, optimize, and reuse AI models.
Tensorflow Hub~\footnote{Available at https://www.tensorflow.org/hub} and PyTorch Hub~\footnote{Available at https://pytorch.org/hub/} enable developers to upload their customized model architecture and pre-trained weights in the cloud database and provide APIs to share public models. However, the AI components shared in PyTorch and Tensorflow Hub have limited their backend which hinders switching the programming frameworks as needed. 

\section{Approach}
In this section, we will present how HPCFair lowered the threshold for diverse scientific communities to adopt AI to their research.
In essence, HPCFair introduced four components to provide scientists with a user-friendly interface.
First, we proposed object converter components to enable programming framework-agnostic implementation.
Then, we introduced AI artifact containerized components, which allow AI artifacts to run independently of the operating system. 
To allow scientists to run AI artifacts effortlessly even without a programming background, we designed a straightforward user query rule and established robust user query processing components.
Additionally, to enable AI artifacts Findable, Accessible, Interoperable, and Reproducible~(FAIR) principles, we  
leveraged an HPC ontology~\citep{liao2021hpc_ontology} to run our proposed platform in HPC clusters.

\subsection{Enable AI artifacts Collaboration}
AI artifacts 
have been developed by different underlying systems, such as different programming languages (Python, C++) and frameworks (Scikt-learn, PyTorch, TensorFlow), and AI artifacts in different frameworks are not transferable. Hence, it raises significant challenges for users inter-operate AI artifacts with distinct underlying backends. For instance, an AI model implemented in C++ is hard to integrate with an AI dataset in Python implementation. 
Domain scientists tend to be challenged to incorporate AI artifacts in their applications, where they have to switch back and forth between different developing backends.

Thanks for recent efforts in ONNX~\cite{bai2019onnx} (a community AI project for building general AI model formats), which uses extensible computation graph models to represent AI models built with different frameworks. Intuitively, with its  framework and platform-independent computational graph representation, AI models developed with different frameworks can be transferred to a general format, and hence, support interoperability between frameworks. 
However, such a great contribution has barely been used by existing HPC-AI tools. 
Therefore
as shown in Figure~\ref{fig:overview} AI artifacts converter, HPCFair developed an online running process that any customized AI model that has been shared, uploaded, and pushed to the HPCFair database would be automatically transferred to ONNX.

\subsection{Containerized AI artifacts}
Since our target users are among different scientific domains and have various hardware environments, we aim to provide solutions for deploying AI artifacts among different platforms. The benefit of an AI model container image can be briefly summarized as follow: first, once the container image is built, it will provide a virtual executing environment for the associate AI model that is independent of local devices. Second, the container image can generalize the model to different software/hardware systems, and save great efforts in environment configurations. Lastly, the container image can be executed easily.

Hence, to improve and reproduce experiments with AI artifacts 
we aim to collect experiments run-time system and supporting metadata configuration information.
Specifically, we leverage MLCube~\cite{kahng2016mlcube} container storing essential runtime experimental configurations and states of AI models.
A containerized object is represented by a configuration file, which contains information on the object's runtime supporting libraries and hyper-parameters. Besides that, uniqueness checks are been performed to guarantee there is no duplicate uploading in the underlying database.

\subsection{User-Friendly Query Rule Design}
As shown in Figure~\ref{fig:overview}, to provide a friendly interactive interface for users, every query made by users would 
initialize the proposed components.
Users may provide configuration files to specify tasks and parameters as needed for their tasks.
In HPCFair, we designed four groups of configuration arguments to conduct main tasks~(store models/datasets, tag-based search, model inference, knowledge transfer/model optimization, load models, and load dataset) provided by HPCFair APIs. Listing~\ref{list:conversion} shows the example configuration for model conversion.
The first configuration arguments group is general arguments, where a user specifies which task to perform, and HPCFair will initialize the corresponding components~(as shown in Listing~\ref{list:conversion} lines 1-3). Then, the user provides the device arguments~(Listing~\ref{list:conversion} lines 5-9), and the user specifies local device information. The next group of configuration arguments is the task arguments~(Listing~\ref{list:conversion} lines 12-16), such as input, working path, etc. Lastly, the output arguments specify where HPCFair exports the output~(Listing~\ref{list:conversion} lines 18-19).

\begin{lstlisting}[caption={Configuration for model conversion},captionpos=b,label={list:conversion}]
general_args:
  task: "conversion"
  backend: ["pt","tf"]

device_args:
  worker_num: 4
  device: "cpu"
  gpu_mapping_file: ''
  gpu_mapping_key: ''

model_args:
  model_name: ["encoder","decoder"]
  model_file: ["./ckpt/encoder.ckpt", "./ckpt/decoder.ckpt"]
  onnx_version: 10

out_args:
  export_file: ["encoder.onnx","decoder.onnx"]
\end{lstlisting}

\subsection{Designed Workflow for FAIR Principles}
Our ultimate goal is to provide scientists with a friendly platform to fetch, share, and apply AI artifacts. 
As shown in Figure~\ref{fig:overview}, we designed an efficient online workflow for HPCFair~\citep{liao2021hpc_ontology}. First, to assist scientists in efficiently finding target AI artifacts~(\textit{Findable}), HPCFair registered and indexed descriptive metadata with corresponding AI artifacts together as a searchable resource. 
The metadata contains rich descriptive information about associated AI artifacts and is assigned a globally unique and persistent identifier, which significantly enhances searchability. 
Second, users can easily access AI artifacts in the HPCFair database through the designed communication protocol~(\textit{Accessible}).
Such communication protocol enables users to share or discover their target AI artifacts efficiently. Additionally, HPCFair also provides authorized credentials for users securely access AI artifacts wherever necessary.
To qualify AI artifacts to 
interoperate among various AI frameworks~(\textit{Interoperable}) at the application level, the object conversion process on the HPCFair server continuously transforms communicated AI models to ONNX format, equipping application users to transform models from one format to another as needed.
Lastly, the scientific community oftentimes interacts among researchers to share and reuse crucial components. HPCFair provides metadata with detailed provenance to reuse the components to build an AI pipeline by plugging the data artifacts~(\textit{Reusable}). The loosely coupled nature of the stored data enables efficient development.

\section{Evaluation}
In this section, we conduct comprehensive evaluations for HPCFair under different practical scenarios and use demos and examples to show the ease of scientists applying AI artifacts by using HPCFair.

\subsection{AI artifacts collaborations}
As AI artifacts are often implemented by diverse frameworks, enabling collaboration among AI artifacts becomes challenging. HPCFair introduces object converter components and provides APIs for a user to allow AI artifacts collaborations.
To assess the HPCFair with a general use case, we experiment with interfacing two AI models implemented with PyTorch and TensorFlow respectively.   
We consider a popular encoder-decoder model structure, given an encoder implemented on PyTorch and a decoder developed by TensorFlow, our goal is to construct an AI model from the given encoder and decoder.

To achieve model collaboration, we first leverage HPCFair APIs to convert target AI artifacts to ONNX formats, then use HPCFair built-in inference API to run the model.
To leverage functional APIs built-in HPCFair, the user provides a straightforward configuration file. In the model collaboration task, we first configure the model conversion task configuration file, as shown in Listing~\ref{list:inference}. As shown in the configuration file, the user specifies the essential AI artifacts information, such as the backend framework, and checkpoint directory. The output file would be saved into the path user defined under $out\_args$.

After the target model has been converted to a uniformed ONNX file, the next step is to run the model. Similarly, HPCFair provides high-level APIs for users to run AI artifacts without programming expertise or knowledge. Listing~\ref{list:inference} shows the inference configuration file.

\begin{lstlisting}[caption={Configuration for model collaboration for inference},captionpos=b,label={list:inference}]
general_args:
  task: "inference"
  tag: "collaboration"
  backend: "onnx"

device_args:
  worker_num: 4
  device: "cpu"
  gpu_mapping_file: ''
  gpu_mapping_key: ''

task_args:
  model_name: ["encoder","decoder"]
  model_file: ["encoder.onnx", "decoder.onnx"]
  onnx_version: 10
  input: "input.txt"

out_args:
  export_file: "out.txt"

\end{lstlisting}

The most exciting part of HPCFair is that it is fairly simple to call the APIs, which usually with one-line codes to finish a task. Listing~\ref{list:api} shows we call HPCFair-provided Python APIs to finish model collaboration tasks based on the configuration files. Model collaboration is a combined task with model conversion and model inference tasks. In the first line, we import the HPCFair python APIs. then in the main function~(lines 3-6), we first create an API object (line 4). Then perform model conversion~(line 5). Lastly, model collaboration ~(line 6).
Taking advantage of the robust high-level APIs, we finish the complex model collaboration task in 3 lines of code.\\
\begin{lstlisting}[language=Python, caption={Call HPCFair APIs },captionpos=b,label={list:api}]
from hpcfair import modelAPI

if __name__ == '__main__':
    api = modelAPI()
    api.conversion(path_to_config)
    api.collaborate(path_to_config)
    api.container(path_to_config)
\end{lstlisting}

\subsection{Inference AI artifacts via HPCFair}
In the AI artifacts inference task, users provide input, and HPCFair runs the target AI artifacts on that input and returns the output. As mentioned before, to support multi-framework and underlying language, HPCFair automatically transfers AI artifacts to ONNX, hence, greatly simplifying the inference process. Inside HPCFair, we build a base container for running ONNX models. The inference examples as shown in Listing~\ref{list:inference} and Listing~\ref{list:api}.








\subsection{Run AI project via HPCFair}
Different from inference AI artifacts, which deal with given inputs, an AI project may involve data processing, training, fine-tuning, and transferring on scaled datasets. 
HPCFair built a running virtual environment for AI projects by containerization. To run the target AI model fetched from HPCFair, users simply provide a configuration file~(as shown in Listing~\ref{list:inference2}). 
HPCFair provides high-level APIs for users to build AI artifacts to their task in one line codes~(Line 7 in Listing~\ref{list:api}).
\begin{lstlisting}[caption={Configuration for running AI project},captionpos=b,label={list:inference2}]
general_args:
  task: "container"
  backend: "mlcube"

device_args:
  device: 'gpu'

task_args:
  work_dir: "project_dir"
  build_file: "path_to_build_file"
  build_tag: "image_name" 
  volume: "/app"
out_args:
  export_file: "out.txt"

\end{lstlisting}
\section{Conclusion}
In conclusion, we proposed a novel model knowledge management system - HPCFair, which enables AI artifacts Findable, Accessible, Interoperable, and Reproducible~(FAIR) principles.   HPCFair provides users with high-level APIs and a friendly interactive interface to fetch, reproduce and retrieve AI artifacts. Most importantly, HPCFair greatly saves the labor cost for scientists to customize AI artifacts to their tasks.

\section*{Acknowledgment}
This research was funded in part by and used resources at the Argonne Leadership Computing Facility, which is a DOE Office of Science User Facility supported under Contract DE-AC02-06CH11357. This
work is also supported by the U.S. Department of Energy,
Office of Science, Advanced Scientific Computing Program
under Award Number DE-SC0021293.

\bibstyle{aaai22}
\bibliography{aaai22}

\begin{thebibliography}{10}
\providecommand{\natexlab}[1]{#1}

\bibitem[{Bai et~al.(2019)Bai, Lu, Zhang et~al.}]{bai2019onnx}
Bai, J.; Lu, F.; Zhang, K.; et~al. 2019.
\newblock ONNX: Open Neural Network Exchange.
\newblock \url{https://github.com/onnx/onnx}.

\bibitem[{Chard et~al.(2019)Chard, Li, Chard, Ward, Babuji, Woodard, Tuecke,
  Blaiszik, Franklin, and Foster}]{chard2019dlhub}
Chard, R.; Li, Z.; Chard, K.; Ward, L.; Babuji, Y.; Woodard, A.; Tuecke, S.;
  Blaiszik, B.; Franklin, M.~J.; and Foster, I. 2019.
\newblock DLHub: Model and data serving for science.
\newblock In \emph{2019 IEEE International Parallel and Distributed Processing
  Symposium (IPDPS)}, 283--292. IEEE.

\bibitem[{Fursin(2021)}]{fursin2021collective}
Fursin, G. 2021.
\newblock Collective knowledge: organizing research projects as a database of
  reusable components and portable workflows with common interfaces.
\newblock \emph{Philosophical Transactions of the Royal Society A}, 379(2197):
  20200211.

\bibitem[{Kahng, Fang, and Chau(2016)}]{kahng2016mlcube}
Kahng, M.; Fang, D.; and Chau, D. H.~P. 2016.
\newblock Visual Exploration of Machine Learning Results Using Data Cube
  Analysis.
\newblock In \emph{Proceedings of the Workshop on Human-In-the-Loop Data
  Analytics}, HILDA '16. New York, NY, USA: Association for Computing
  Machinery.
\newblock ISBN 9781450342070.

\bibitem[{Kurtzer, Sochat, and Bauer(2017)}]{kurtzer2013singularity}
Kurtzer, G.~M.; Sochat, V.; and Bauer, M.~W. 2017.
\newblock Singularity: Scientific containers for mobility of compute.
\newblock \emph{PLOS ONE}, 12(5): e0177459.

\bibitem[{Liao et~al.(2021)Liao, Lin, Verma, Vanderbruggen, Emani, Nan, and
  Shen}]{liao2021hpc_ontology}
Liao, C.; Lin, P.-H.; Verma, G.; Vanderbruggen, T.; Emani, M.; Nan, Z.; and
  Shen, X. 2021.
\newblock HPC Ontology: Towards a Unified Ontology for Managing Training
  Datasets and AI Models for High-Performance Computing.
\newblock In \emph{2021 IEEE/ACM Workshop on Machine Learning in High
  Performance Computing Environments (MLHPC)}, 69--80. IEEE.

\bibitem[{Merkel(2014)}]{merkel2014docker}
Merkel, D. 2014.
\newblock Docker: lightweight linux containers for consistent development and
  deployment.
\newblock \emph{Linux journal}, 2014(239): 2.

\bibitem[{Nan et~al.(2021)Nan, Guan, Shen, and Liao}]{nan2021deep}
Nan, Z.; Guan, H.; Shen, X.; and Liao, C. 2021.
\newblock Deep nlp-based co-evolvement for synthesizing code analysis from
  natural language.
\newblock In \emph{Proceedings of the 30th ACM SIGPLAN International Conference
  on Compiler Construction}, 141--152.

\bibitem[{Verma et~al.(2021)Verma, Emani, Liao, Lin, Vanderbruggen, Shen, and
  Chapman}]{verma2021hpcfair}
Verma, G.; Emani, M.; Liao, C.; Lin, P.-H.; Vanderbruggen, T.; Shen, X.; and
  Chapman, B. 2021.
\newblock HPCFAIR: Enabling FAIR AI for HPC Applications.
\newblock In \emph{2021 IEEE/ACM Workshop on Machine Learning in High
  Performance Computing Environments (MLHPC)}, 58--68. IEEE.

\bibitem[{Wolf et~al.(2019)Wolf, Debut, Sanh, Chaumond, Delangue, Moi, Cistac,
  Rault, Louf, Funtowicz et~al.}]{wolf2019huggingface}
Wolf, T.; Debut, L.; Sanh, V.; Chaumond, J.; Delangue, C.; Moi, A.; Cistac, P.;
  Rault, T.; Louf, R.; Funtowicz, M.; et~al. 2019.
\newblock Huggingface's transformers: State-of-the-art natural language
  processing.
\newblock \emph{arXiv preprint arXiv:1910.03771}.

\end{thebibliography}

\end{document}